\documentclass[a4paper,fleqn,usenatbib]{mnras}

\usepackage{mathptmx}
\usepackage{txfonts}

\usepackage[T1]{fontenc}
\usepackage{ae,aecompl}
\usepackage{pdflscape}

\usepackage{graphicx}	
\usepackage{multicol, wrapfig}

\title[VVV-WIT-01]{VVV-WIT-01: highly obscured classical nova or protostellar collision?}

\author[P. W. Lucas et al.]{
P. W. Lucas,$^{1}$\thanks{E-mail: p.w.lucas@herts.ac.uk (PWL)}
D. Minniti$^{2,3,4}$, A. Kamble$^{5}$, D. L. Kaplan$^{6}$, N. Cross$^{7}$,  I. Dekany$^{8}$, 
\newauthor V.D. Ivanov$^{9}$, R. Kurtev$^{2,10}$, R.K. Saito$^{11}$, L.C. Smith$^{12}$, M. Catelan$^{2,13}$,
\newauthor N. Masetti$^{14,3}$, I. Toledo$^{15}$, M. Hempel$^{3}$, M.A. Thompson$^{1}$,  C. Contreras Pe\~{n}a$^{16}$,
\newauthor J. Forbrich$^{1}$, M. Krause$^{1}$, J. Dale$^{1}$,  J. Borissova$^{2,10}$, J. Emerson$^{17}$\\
$^{1}$Centre for Astrophysics, University of Hertfordshire, College Lane, Hatfield, AL10 9AB, UK\\
$^{2}$Millennium Institute of Astrophysics, Av. Vicuna Mackenna 4860, 782-0436, Macul, Santiago, Chile\\
$^{3}$Departamento de Ciencias F\'{i}sicas, Universidad Andr\'{e}s Bello, Fern\'{a}ndez Concha 700, Las Condes, Santiago, Chile\\
$^{4}$Vatican Observatory, V00120 Vatican City State, Italy\\
$^{5}$Harvard-Smithsonian Center for Astrophysics, 60 Garden St., Cambridge, MA 02138, USA\\
$^{6}$Center for Gravitation, Cosmology and Astrophysics, University of Wisconsin-Milwaukee, P.O. Box 413, Milwaukee, WI 53201, USA\\
$^{7}$Wide-Field Astronomy Unit, Institute for Astronomy, University of Edinburgh, Royal Observatory, Blackford Hill, Edinburgh EH9 3HJ, UK\\ 
$^{8}$Astronomisches Rechen-Institut, Zentrum f\"{u}r Astronomie der Universit\"{a}t Heidelberg, M\"{o}nchhofstr. 12-14, D-69120 Heidelberg, Germany\\
$^{9}$European Southern Observatory, Karl-Schwarszchild-Str 2, D-85748 Garching bei Muenchen, Germany\\
$^{10}$Instituto de F\'{i}sica y Astronom\'{i}a, Universidad de Valpara\'{i}so, ave. Gran Breta\~{n}a, 1111, Casilla 5030, Valpara\'{i}so, Chile\\
$^{11}$Departamento de F\'{i}sica, Universidade Federal de Santa Catarina, Trindade 88040-900, Florianopolis, SC, Brazil\\
$^{12}$Institute of Astronomy, University of Cambridge, Madingley Road, Cambridge, CB3 0HA, UK\\
$^{13}$Instituto de Astrofisica, Pontificia Universidad Catolica de Chile, Av. Vicuna Mackenna 4860, 7820436 Macul, Santiago, Chile\\
$^{14}$INAF - Osservatorio di Astrofisica e Scienza dello Spazio, via Gobetti 93/3, I-40129 Bologna, Italy\\
$^{15}$ALMA Observatory, Alonso de Cordova 3107, Vitacura, Santiago, Chile\\
$^{16}$School of Physics, University of Exeter, Stocker Road, Exeter, EX4 4QL, UK\\
$^{17}$Astronomy Unit, School of Physics and Astronomy, Queen Mary University of London, Mile End Road, London, E1 4NS, UK}
\date{Accepted XXX. Received YYY; in original form ZZZ}
\pubyear{2019}
\begin{document}
\label{firstpage}
\pagerange{\pageref{firstpage}--\pageref{lastpage}}
\maketitle
\begin{abstract}
A search of the first Data Release of the VISTA Variables in the Via Lactea (VVV) Survey discovered the exceptionally red transient VVV-WIT-01 ($H-K_s=5.2$). It peaked before March 2010, then faded by $\sim$9.5 mag over the following two years. The 1.6--22~$\mu$m spectral energy distribution in March 2010 was well fit by a highly obscured black body with $T \sim 1000~K$ and $A_{K_s} \sim 6.6$~mag. The source is projected against the Infrared Dark Cloud (IRDC) SDC G331.062$-$0.294. The chance projection probability is small for any single event ($p \approx 0.01$ to 0.02) which suggests a physical association, e.g. a collision between low mass protostars. However, black body emission at $T \sim 1000~K$ is common in classical novae (especially CO novae) at the infrared peak in the light curve, due to condensation of dust $\sim$30--60 days after the explosion. Radio follow up with the Australia Telescope Compact Array (ATCA) detected a fading continuum source with properties consistent with a classical nova but probably inconsistent with colliding protostars. Considering all VVV transients that could have been projected against a catalogued IRDC raises the probability of a chance association to $p=0.13$ to 0.24. After weighing several options, it appears likely that VVV-WIT-01 was a classical nova event located behind an IRDC.
\end{abstract}


\section{Introduction}

\begin{figure*}
	\includegraphics[width=\textwidth]{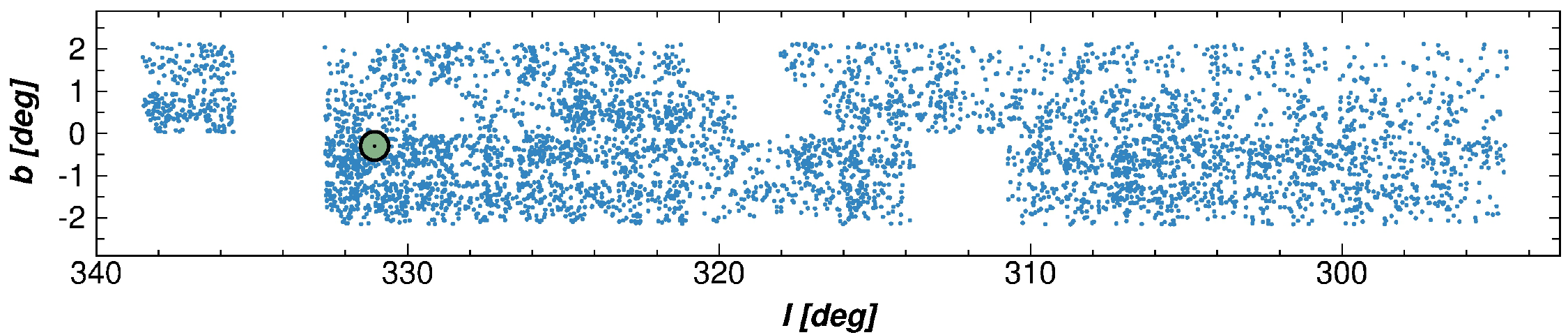}
    \caption{The VVV DR1 survey area in the southern Galactic plane, showing the locations of all the candidate high-amplitude variable stars returned by the SQL query. The total area explored covers about 150~deg$^2$ within $\pm 2^{\circ}$ of the Galactic plane. The green circle represents the position of VVV-WIT-01.}
    \label{fig:wit01_f1}
\end{figure*}

\begin{figure}
	\includegraphics[width=\columnwidth]{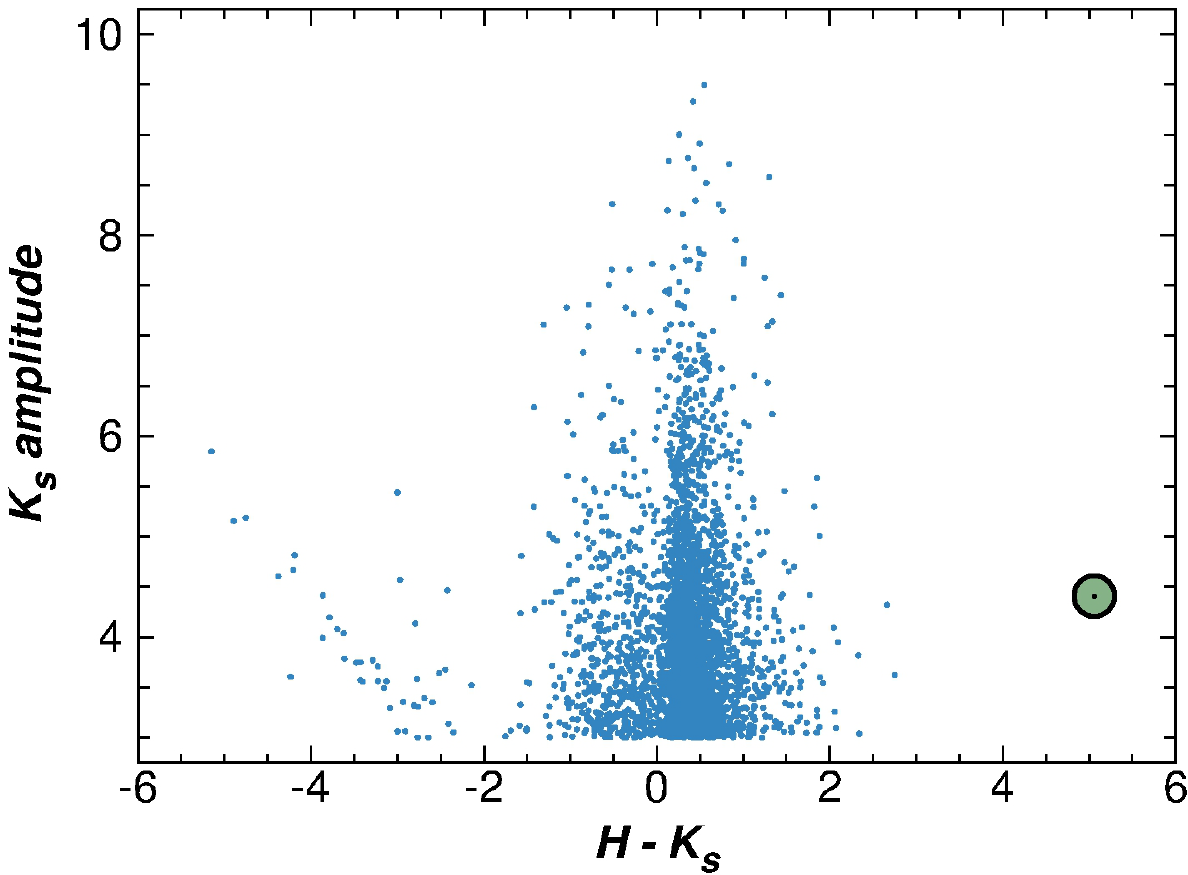}
    \caption{$K_s$ amplitudes ($K_{s, max}-K_{s, min}$) vs. $H-K_s$ colour for candidate high amplitude variable stars returned by the SQL query of VVV DR1. The    green circle indicates VVV-WIT-01.}
    \label{fig:wit01_f2}
\end{figure}

Very high amplitude near infrared variable sources detected in the Milky Way are typically cataclysmic variables (CVs) \citep[e.g.][]{saito15, saito16, de19} or Young Stellar Objects (YSOs) undergoing episodic accretion, variable extinction or both \citep[e.g.][]{kospal13, cp17a, lucas17}.  The VVV survey \citep{minniti10, saito12} is the first panoramic near infrared time domain survey of the Milky Way, offering the chance to improve our knowledge of infrared variability and to discover new types of variable star. Studies of VVV transients may assist extragalactic transient work, where events such as Luminous Red Novae (LRN) have uncertain physical origin \citep[e.g.][]{kasliwal11, pastorello19}. In the Milky Way, the remnants of past explosive events can readily be detected, providing examples of events that have not been seen in our galaxy in modern times. For example the Becklin-Neugebauer explosion in the Orion Molecular Cloud 1 (OMC-1), likely the consequence of a protostellar collision, occurred $\sim$500 years ago and the optically obscured supernova remnant G1.9$+$0.3 appears to be only $\sim$100 years old \citep{reynolds08}.
In this paper we report the discovery and analysis of VVV-WIT-01, an extremely red transient that was first observed by the VVV survey in March 2010. 

\section{VVV-WIT-01 discovery data}
\label{sec:2}

Since 2010, the VVV survey has mapped and monitored the Galactic bulge and the adjacent southern plane in the near infrared with VIRCAM (VISTA InfraRed CAMera) mounted on the 4.1~m wide-field Visible and Infrared Survey Telescope for Astronomy \citep[VISTA,][]{sutherland15} at ESO Paranal Observatory. There are now typically between 75 and 100 epochs of data in the $K_s$ bandpass for each field, across a total area of 560 deg$^2$. From 2016, the survey area was extended to cover new fields (the rest of the southern Galactic plane at longitudes $l>230^{\circ}$ and $l<20^{\circ}$, along with the outer bulge \citep{minniti16} leading to less frequent observations of the original VVV fields.
Initial data reduction and archival merging are carried out using the pipelines at the Cambridge Astronomical Survey Unit \citep[CASU,][]{irwin04} and VISTA Science Archive (VSA) in Edinburgh \citep{cross12}. The VVV photometry presented here are profile fitting photometry, performed with an updated version of {\sc DoPhot} \citep{schechter93, alonso12} for all data in the original VVV fields (Smith et al., in prep). The absolute calibration is based on the VVV photometric catalogue of \citet{alonso18} which is in turn derived from the CASU v1.3 VISTA pipeline. The VVV Survey produces near-IR multi-colour photometry in five passbands: $Z$ (0.88~$\mu$m), $Y$ (1.02~$\mu$m), $J$ (1.25~$\mu$m), $H$ (1.65~$\mu$m), and $K_s$ (2.15~$\mu$m), as well as multi-epoch photometry in $K_s$.

The VVV 1st Data Release (VVV DR1), an SQL-based relational database available at the VSA, contains the VVV data taken up to 30 Sep 2010. It includes about 58 million sources detected at 4 epochs or more in a 150~deg$^2$ area at Galactic latitudes $b < \pm2^{\circ}$ (see Fig. \ref{fig:wit01_f1}).
VVV DR1 was created in early 2012 and an SQL query was very soon run to search for transients and variable sources with $>$3 magnitudes of variation. The query selected candidates having at least 4 epochs of observation, images in both the $H$ and $K_s$ bandpasses and a median $K_s$ magnitude at least two magnitudes brighter than the expected limiting magnitude for each field. There were a total of $\sim$5000 candidates, the great majority of them false positives with unremarkable colours  ($H-K_s<2$) but one extremely red candidate with ($H-K_s = 5.2$) stood out in the colour vs. amplitude diagram (see Fig. \ref{fig:wit01_f2}). After passing visual inspection, it was classified as the first ``WIT" source (an acronym for ``What is This?") deemed worthy of further study. It was later renamed as VVV-WIT-01. (False positives were due to multiple causes such as saturated stars, blends of two or more stars or cosmic rays.)

\begin{figure}
	\hspace{4mm} \includegraphics[width=0.87\columnwidth]{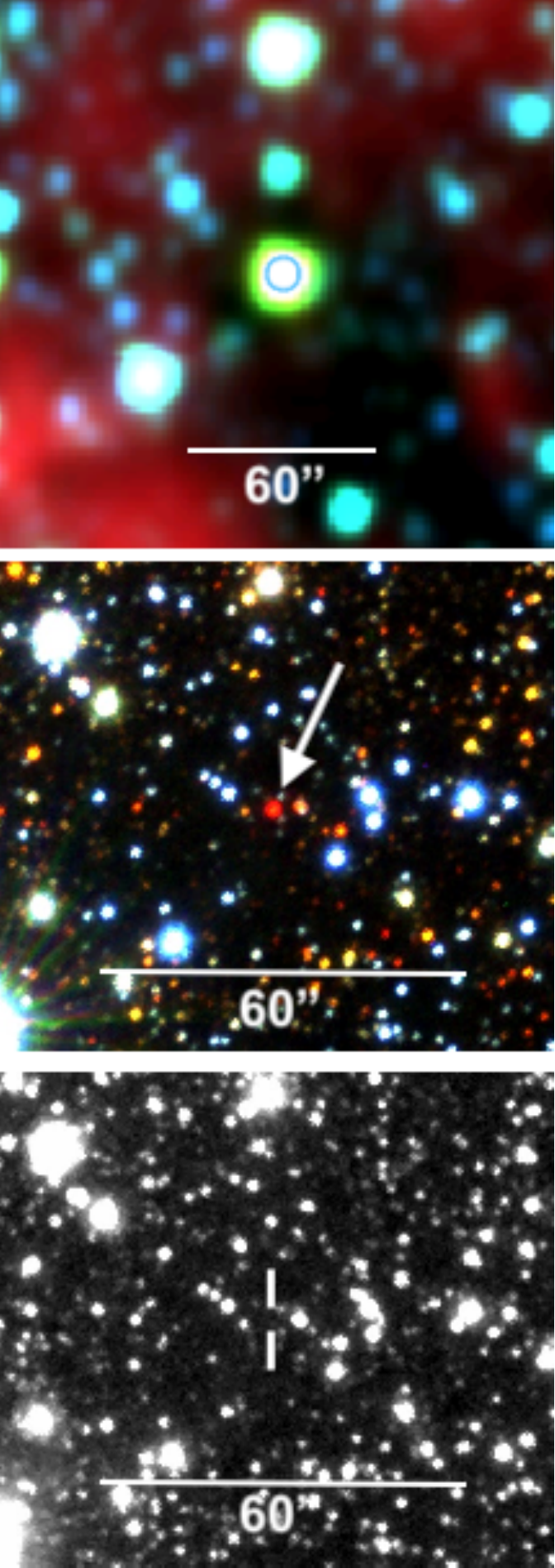}
    \caption{Finder images (equatorially oriented). ({\it Top}:) $3\times3\arcmin$ WISE 3-colour image showing VVV-WIT-01 in 2010, marked with a blue circle (blue: 3.3~$\mu$m, green: 4.6~$\mu$m, red: 12~$\mu$m). The green colour is due to high extinction and emission from warm circumstellar matter. ({\it Middle:}) 
    $90\times80\arcsec$ VVV 3-colour image (blue: $J$, green $H$, red: $K_s$) from March 2010. VVV-WIT-01 is the very red star at the centre, with additional red 
    YSO candidates adjacent. ({\it Bottom:}) $K_s$ stacked image made from all VVV images of the field in 2011. The transient was no longer detected.} 
    \label{fig:wit01_f3}
\end{figure}

VVV-WIT-01 is the only high amplitude variable with such a red $H-K_s$ colour amongst 58 million sources in DR1 with 4 or more epochs of photometry \citep{minniti12}. Moreover, additional searches of a database of 2010--2018 light curves for all sources in the original 560~deg$^2$ VVV area have found no additional transients with $H-K_s > 3$ (see section \ref{sec:addtran}).
It was at $K_s \approx 12.2$ when first detected in March 2010 but faded to $K_s \sim 16.7$ by July of that year and was then undetected in VVV images taken in every subsequent year up to 2018, to a typical 3$\sigma$ upper limit of $K_s = 18.85$ (using annual stacks of all the relevant images taken each year).  The source was not detected in the VVV $Z$, $Y$ and $J$ images taken in 2010 and again in 2015, to 3$\sigma$ limits of 20.5, 20.5 and 20.0 mag in $Z$, $Y$ and $J$ respectively. Finder images showing the location of the source are presented in Fig. \ref{fig:wit01_f3} and the time series of profile fitting photometry is given in Table 1.

\begin{figure*}
	\includegraphics[width=0.85\textwidth]{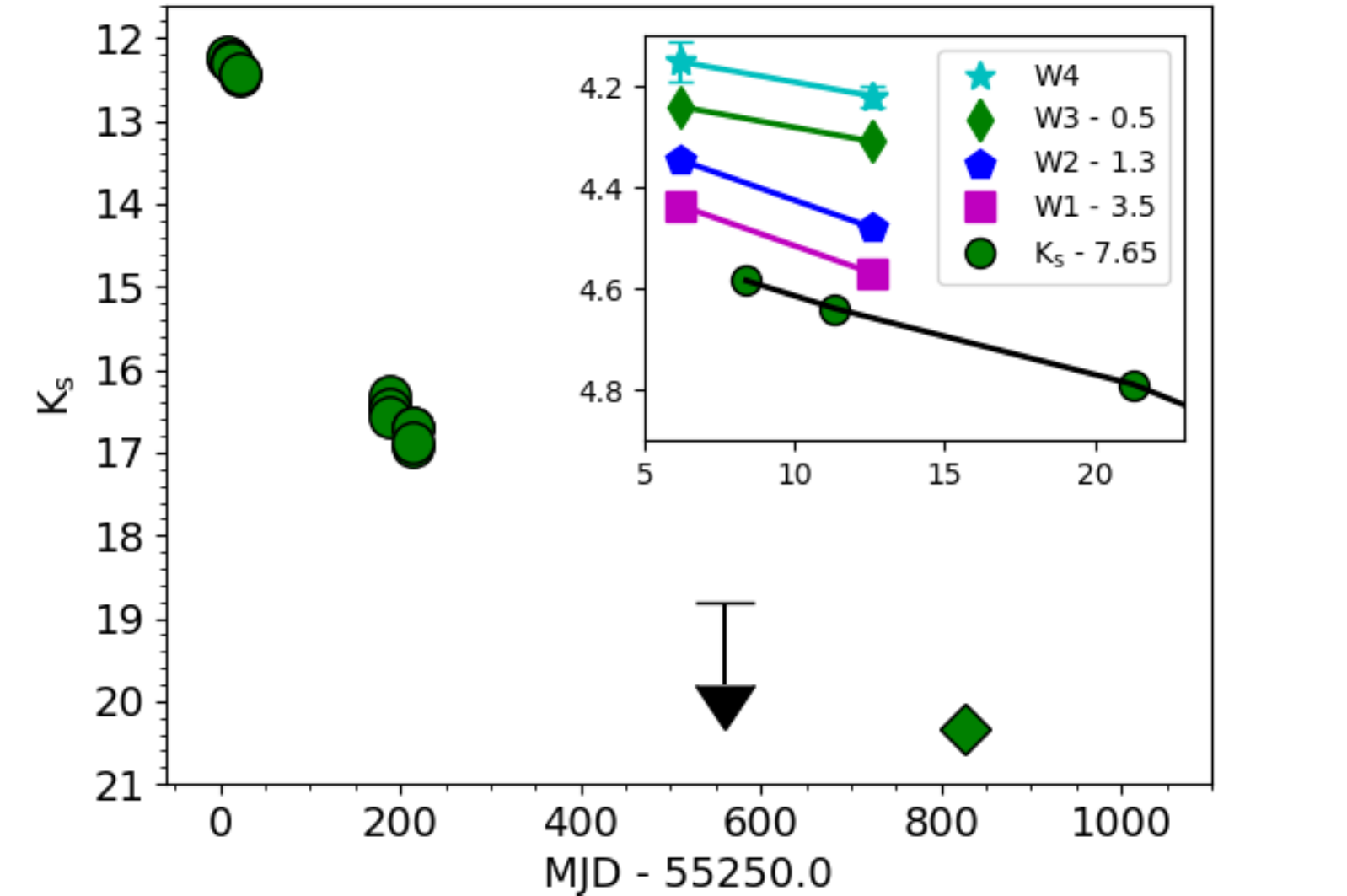}
    \caption{Observed $K_s$ light curve for VVV-WIT-01. We show the VVV Survey observations from 2010 (green circles), until 2011 (upper limit), and the ISAAC observation from May 2012 (green diamond). The insert shows the contemporaneous VVV and WISE observations from early 2010, where we have averaged data within each day for clarity. The error bars are typically 
smaller than the point sizes.}
    \label{fig:wit01_f4}
\end{figure*}

\begin{figure*}
	\includegraphics[width=0.85\textwidth]{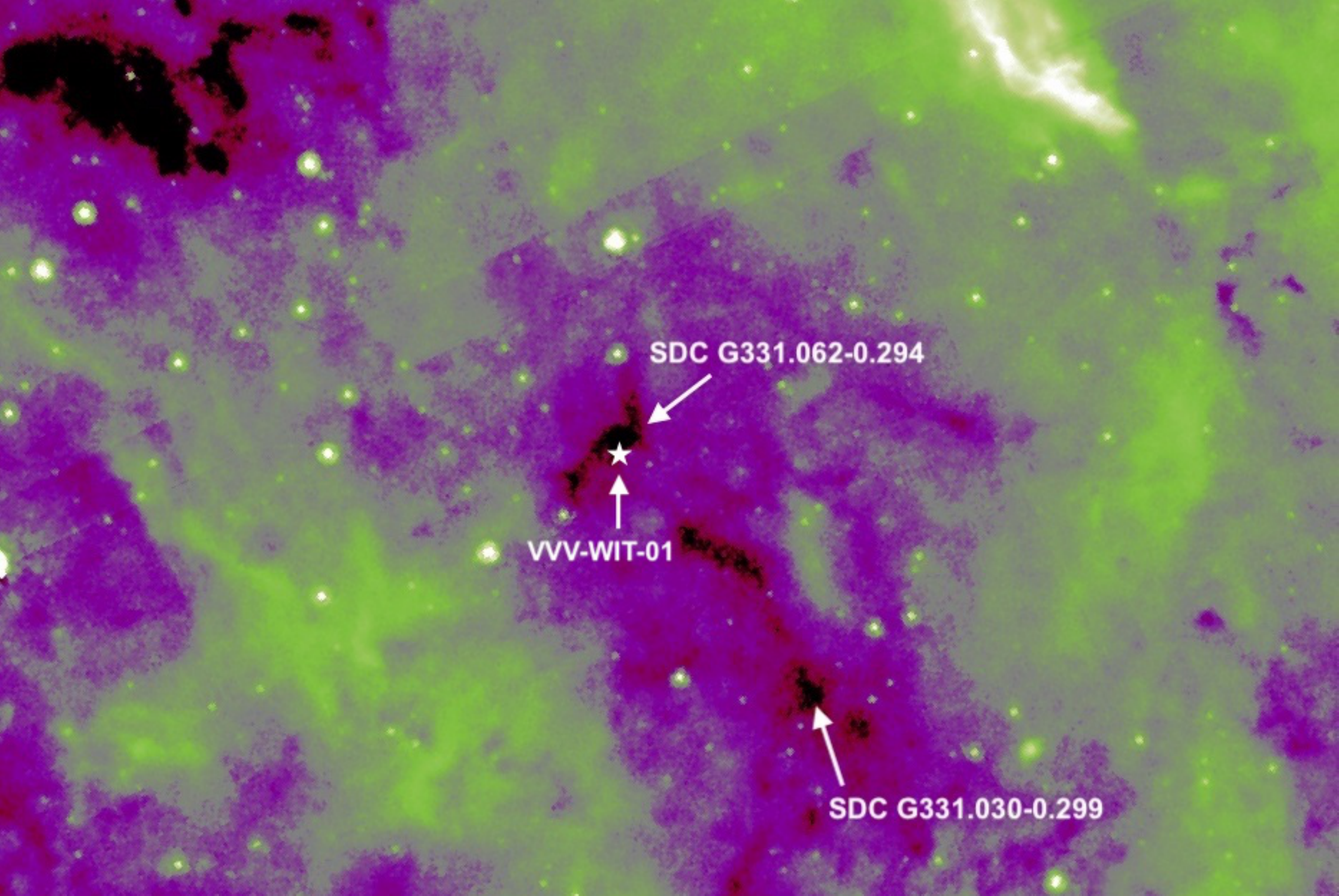}
    \caption{{\it Spitzer}/GLIMPSE 8~$\mu$m  $8 \times 6$\arcmin~false colour image of the IRDCs in the VVV-WIT-01 field, with equatorial orientation. This 
    image was taken in 2004, before the transient event (marked with a white star) in SDC~G331.062$-$0.294. We see that this IRDC is part of a larger dark region that 
    includes SDC G331.030$-$0.299. The bright diffuse 8~$\mu$m background emission (green or white areas) is typical of the mid-plane of the inner 
    Milky Way. The white area at the top right is the HII region IRAS~16067-5145 at $d \approx 7.4$~kpc, see text.}
    \label{fig:wit01_f5}
\end{figure*}

\begin{table}
	\begin{centering}
	\caption{VVV Photometry of VVV-WIT-01.} 
	\label{tab:table1}
	\begin{tabular}{rccc} \hline
MJD-2455250.0 & Filter & Magnitude & $\sigma$\\ \hline
8.381478 & $K_s$ & 12.24 & 0.02\\
8.382969 & $K_s$ & 12.23 & 0.02 \\
11.339142 & $K_s$ & 12.28 & 0.02 \\
11.339855 & $K_s$ & 12.30 & 0.02 \\
21.294046 & $K_s$ & 12.44 & 0.02 \\
21.295505 & $K_s$ & 12.43 & 0.02 \\
187.054863 & $K_s$ & 16.42 & 0.06 \\
187.055789 & $K_s$ & 16.32 & 0.05 \\
189.022168 & $K_s$ & 16.47 & 0.05 \\
189.022994 & $K_s$ & 16.57 & 0.06 \\
213.004979 & $K_s$ & 16.92 & 0.10 \\
213.005872 & $K_s$ & 16.69 & 0.07 \\
214.021768 & $K_s$ & 16.70 & 0.09 \\
214.022686 & $K_s$ & 16.87 & 0.11 \\
8.376614     & $H$ &   17.50 & 0.04 \\
8.378080     & $H$ &   17.41 & 0.04 \\
21.289327   & $H$ &   17.60 & 0.02 \\
21.290754   & $H$ &   17.62 & 0.05 \\  \hline
	\end{tabular}
	\end{centering}\\
	Note: the quoted errors are those given by {\sc DoPhot}, with a floor of 0.02~mag imposed as the estimated 
	calibration uncertainty.
\end{table}

\begin{table*}
	\begin{centering}
	\caption{AllWISE Multi-epoch Photometry of VVV-WIT-01.} 
	\label{tab:table2}
	\begin{tabular}{lcccccccc} \hline
MJD-55250 & $W1$ & $\sigma_{W1}$ & $W2$ & $\sigma_{W2}$ & $W3$ & $\sigma_{W3}$ & $W4$ & $\sigma_{W4}$ \\ \hline
5.564351 & 7.918 & 0.018 & 5.592 & 0.018 & 4.764 & 0.028 & 4.295 & 0.149 \\
5.696656 & 7.908 & 0.015 & 5.669 & 0.022 & 4.753 & 0.027 & 4.117 & 0.111 \\
5.828960 & 7.929 & 0.016 & 5.685 & 0.022 & 4.746 & 0.024 & 4.066 & 0.116 \\
5.961264 & 7.922 & 0.015 & 5.775 & 0.019 & 4.747 & 0.031 & 4.115 & 0.164 \\
6.027479 & 7.915 & 0.017 & 5.63 & 0.016 & 4.795 & 0.03 & 4.211 & 0.193 \\
6.093695 & 7.934 & 0.017 & 5.603 & 0.024 & 4.743 & 0.024 & 4.025 & 0.137 \\
6.159783 & 7.922 & 0.013 & 5.667 & 0.021 & 4.761 & 0.023 & 4.157 & 0.161 \\
6.225999 & 7.975 & 0.016 & 5.672 & 0.029 & 4.696 & 0.023 & 4.137 & 0.167 \\
6.292087 & 7.945 & 0.014 & 5.621 & 0.017 & 4.745 & 0.027 & 4.153 & 0.16 \\
6.424391 & 7.974 & 0.017 & 5.546 & 0.019 & 4.722 & 0.029 & 4.113 & 0.16 \\
6.424519 & 7.939 & 0.016 & 5.575 & 0.015 & 4.75 & 0.033 & 4.253 & 0.165 \\
6.556823 & 7.945 & 0.015 & 5.742 & 0.028 & 4.718 & 0.023 & 4.14 & 0.085 \\
6.689127 & 7.942 & 0.015 & 5.627 & 0.017 & 4.698 & 0.027 & 4.196 & 0.195 \\
12.57761 & 8.078 & 0.018 & 5.805 & 0.016 & 4.806 & 0.026 & 4.2 & 0.166 \\
12.57773 & 8.057 & 0.021 & 5.754 & 0.021 & 4.819 & 0.026 & 4.243 & 0.171 \\ \hline	
	\end{tabular}
	\end{centering}\\
\end{table*}

The observed light curve (see Fig. \ref{fig:wit01_f4}) shows a rate of decline similar to that of a core collapse supernova  \citep{meikle00, mannucci03, leibundgut03} or perhaps a classical nova, given that the latter have widely varying decline timescales (see section \ref{sec:coll}).
A search of the SIMBAD database revealed that the object is projected within SDC G331.062$-$0.294, a small  Infrared Dark Cloud (IRDC) from the catalogue of \citet{peretto09} with an equivalent-area radius of 17\arcsec. This IRDC catalogue in fact lists two IRDCs within two arcminutes (the other being SDC G331.030$-$0.299) and inspection of the {\it Spitzer}/GLIMPSE 8~$\mu$m image for the region shows that these two appear to be part
of the same dark region $\sim$3\arcmin~in extent (see Fig. \ref{fig:wit01_f5}) that is seen in projection against the inner part of a large bubble of bright mid-infrared 
emission designated as MWP1G331057$-$002239 \citep{simpson12}. This bubble is coincident with the HII region IRAS 16067$-$5145, for which \citet{jones12} 
give a kinematic distance of 7.4$\pm$1.1~kpc.
IRDCs are the densest portions of molecular clouds and the probability of chance association with such an object is only $\sim$1\%  (see section \ref{sec:discussion}) for any single unrelated VVV transient within the southern half of the Spitzer/GLIMPSE~I survey area (where the \citet{peretto09} catalogue is defined). We must therefore consider the case for a pre-main sequence origin for the event. We should also keep in mind that many transients have been detected in VVV so the chance of observing at least one transient in an IRDC over the course of the survey is rather higher (discussed later in section \ref{sec:addtran}). While a location within or behind an IRDC may appear to explain the red colour simply as a consequence of very high extinction, this alone cannot explain the slope of the spectral energy distribution (SED) in the mid-infrared (see section \ref{sec:irop}).

IRDCs are initially detected simply as dark regions against the bright mid-infrared background of the Galactic plane. Small IRDCs from the \citet{peretto09} catalogue can be false positive detections caused by dips in the background emission so confirmation is typically sought via detection of far-infrared emission from cold dust in the putative dense cloud. \citet{peretto16} performed automated checks on their 2009 catalogue using the 250~$\mu$m {\it Herschel}/HiGal images \citep{molinari10}
to statistically assess the fraction of false positives and false negatives. SDC~G331.062$-$0.294 passed only one of the two principal checks: far infrared emission was detected but below the 3$\sigma$ threshold, because the measured noise level was inflated by real variation on small spatial scales. However, the adjacent cloud SDC G331.030$-$0.299 passed both checks.
Visual inspection of the HiGal 250~$\mu$m image clearly shows emission at the two expected locations (brightest in between them in fact) and the surrounding bright and highly structured emission from the [SPK2012] MWP1G331057$-$002239 bubble dominates the image. In view of the caveats described by \citet{peretto16} for their automated analysis and after visual comparison with other small IRDCs in the same HiGal image that pass or fail the automated tests, we do not regard this non-confirmation of SDC~G331.062$-$0.294 as significant. Given that high infrared extinction is required to fit the SED of VVV-WIT-01 (see section \ref{sec:irop}) and YSO candidates are present (see section \ref{sec:alma} and Appendix A) it seems very probable that SDC~G331.062$-$0.294 is a bona fide IRDC located in front of the HII region IRAS 16067-5145.

In summary, VVV-WIT-01 is a very red object ($H-K_s$ = 5.22 $\pm$~0.05~mag), with mean $K_s$ = 12.3 mag in March 2010 (Fig. \ref{fig:wit01_f4}). We have measured an accurate absolute position (uncertainty $\sim$6~milliarcsec) based on the VVV profile fitting photometry catalogues, astrometrically transformed to the Gaia DR2 reference frame \citep{lindegren18} using numerous stars in the vicinity of the transient (Smith et al., in prep.). The J2000.0 equatorial coordinates are: RA = 16h 10m 53.4293s, DEC = -51d 55m 32.439s (J2000), and the Galactic coordinates are: $l = 331.061347^{\circ}$, $b=0.295905^{\circ}$.

\begin{figure}
	\includegraphics[width=\columnwidth]{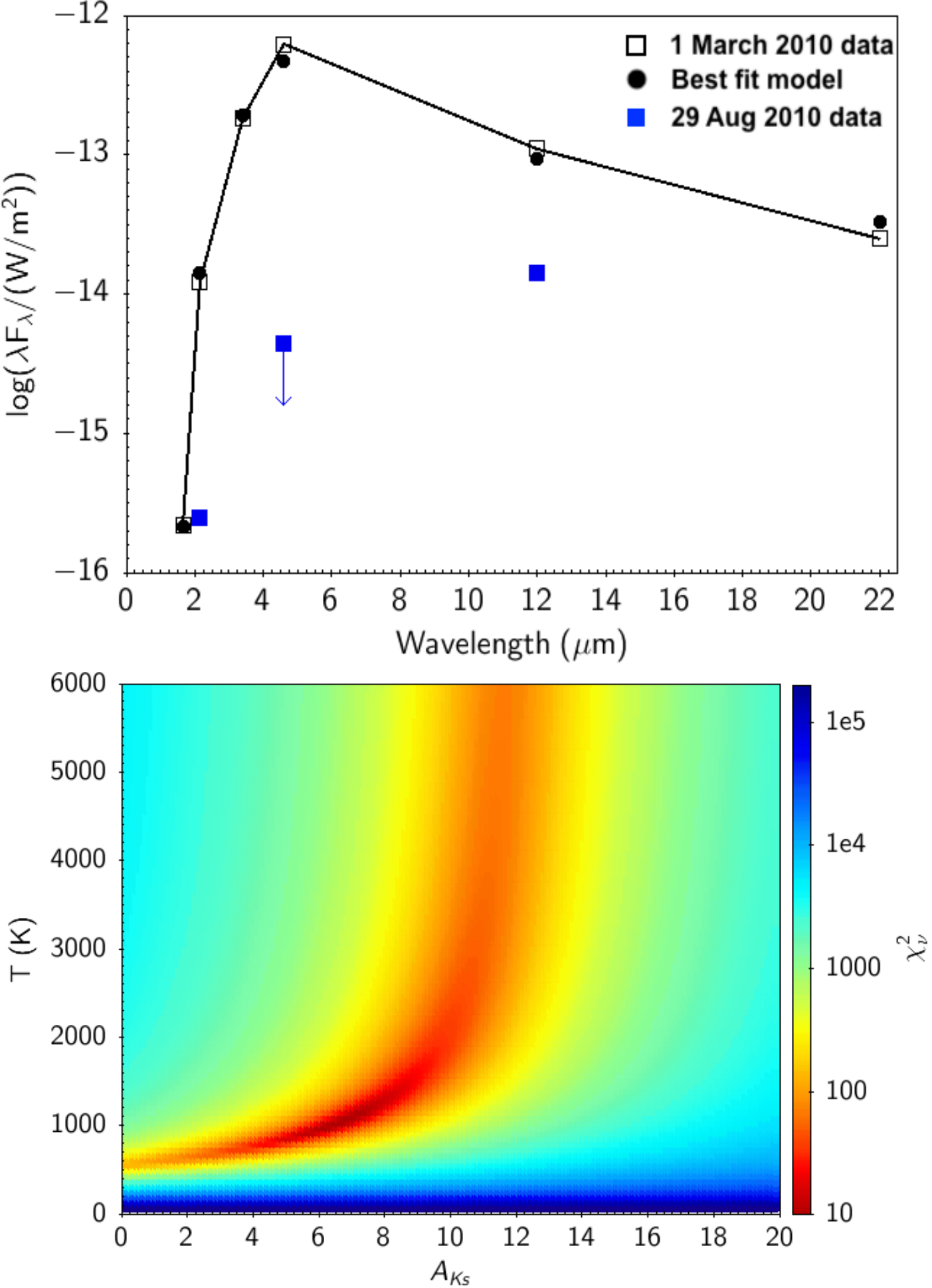}
    \caption{Fit to the SED of VVV-WIT-01. {\it (upper panel:)} The SED around 1 March 2010 is well fit by a black body model with high extinction 
    ($T=1020$~K, $A_{K_s}=6.6$). Blue points show the SED on 29 August 2010, including the upper limit at 4.6~$\mu$m. {\it (lower panel:)}  
    A plot of $\chi^2_\nu$ as a function of extinction and temperature.}
    \label{fig:wit01_f6}
\end{figure}

\section{Multiwaveband follow up and archival observations} 

\subsection{Infrared and optical data}
\label{sec:irop}

We searched various archival optical-infrared databases for this source. The source is absent in the optical plates taken with the UK Schmidt Telescope, available at
the SuperCosmos Sky Survey (\url{www-wfau.roe.ac.uk/sss/}). There were no detections in the following additional image datasets: $JHK_s$ data taken on 1999 July 15 by the Two Micron All Sky Survey \citep{skrutskie06}, $IJK$ data taken on 1996 April 16 by the Deep Near Infrared Survey of the Southern Sky \citep{epchtein99}, $riH\alpha$ data  taken on 2015 June 8 by VPHAS$+$ \citep{drew14}, 
3.6--8.0~$\mu$m mid-infared data taken on 2004 April 2 by {\it Spitzer}/GLIMPSE \citep{benjamin03}, a 24~$\mu$m image taken on 2006 April 13 by {\it Spitzer}/MIPSGAL  \citep{rieke04, carey09} and {\it Midcourse Space Experiment} 8~$\mu$m and 21~$\mu$m images taken in 1996--1997 \citep{egan98}.

However, the object is present and very bright in the Wide-field Infrared Survey Explorer (WISE) mid-infrared dataset (with several images between 2010 February 28 and 2010 March 7, fortuitously overlapping with the dates of the first VVV images). The WISE photometry from the AllWISE Multi-Epoch Photometry Database  is given
 in Table 2.  The tabulated $W2$ data in fact require a positive correction for saturation of 0.07 mag \citep{cutri12}; this applies in a similar fashion for WISE Allsky and AllWISE data taken during the cryogenic portion of the mission. After making this correction, the mean magnitudes are $W1$(3.3~$\mu$m) = 7.95 $\pm$ 0.01, $W2$(4.6~$\mu$m) = 5.73 $\pm$ 0.05, $W3$(12~$\mu$m) = 4.75 $\pm$ 0.01, and $W4$(22~$\mu$m) = 4.16 $\pm$ 0.02 mag. The brightness of VVV-WIT-01 steadily declined in March 2010 (see Fig. \ref{fig:wit01_f4}) matching the slope of the near-simultaneous VVV observations in $K_s$. 
 
After March 2010, the source was no longer detected in the sensitive WISE $W1$ or $W2$ passbands (we inspected the deeper unWISE W2 stacked image constructed from data taken on 2010 August 28 to August 31 \citep{meisner18} but the location is severely blended with 2 adjacent stars in the $\sim$6\arcsec~WISE beam, preventing detection of stars with $W2 \ge 11$~mag). However, the stacked $W3$ image taken at that time shows a clear but uncatalogued detection of a point source at the position of the transient. There are no blending issues in this waveband because fewer stars are detected. Moreover, this source is not present in the GLIMPSE 8~$\mu$m image taken in 2004 (see Fig. \ref{fig:wit01_f5}) so we can be confident that it is VVV-WIT-01. We suspect that the AllWISE pipeline did not
detect it because the sky background level is measured with a coarse spatial grid so the flux peak within the small IRDC did not stand out sufficiently above the (overestimated) background level. We performed aperture photometry on the $W3$ FITS image obtained from the NASA Infrared Science Archive with {\small IRAF/DAOPHOT}, bootstrapping the zero point using several nearby sources in the AllWISE source catalogue. We find $W3=6.78 \pm 0.14$ mag on 2010 August 29 
(the mid-point date of the image stack), the measurement error being dominated by the uncertain background level within the IRDC. This compares with $K_s = 16.45$ on the same date (averaging VVV data from August 28 and August 30), indicating that the transient had faded by 4.2 mag at 2.15~$\mu$m but only 2.0 mag at 12~$\mu$m. N.B. the August 2010 data were from the "3-Band Cryo" phase of the WISE mission so no data were taken in $W4$.

The near simultaneity of the VVV and WISE measurements in March 2010 allows us to define an SED, see Fig. \ref{fig:wit01_f6} (upper panel). We attempted to fit the SED using a reddened, uniform-temperature black body, i.e. the model is $\log_{10}(\lambda B_{\lambda}(T) 10^{-0.4A(\lambda)})$. We constructed a library of black body SEDs with a range of temperatures and extinctions (60<T<6000~K and $0<A_{K_s}<20$) and located the area of parameter space with the minimum value of the reduced $\chi^2$ parameter, $\chi^2_\nu$, computed with five degrees of freedom. Each model was scaled to the same average flux as the data, measured as the average
of log($\lambda F_{\lambda}$) at the six wavelengths. To relate extinction in $K_s$ and the WISE passbands, we adopted the mid-infrared extinction law defined by \citet{koenig14} for fields with high extinction;  separately, the \citet{cardelli89} extinction law was used to relate the extinction in $H$ and $K_s$\footnote{We also considered the modified near infrared extinction law of \citet{alonso17}, derived for VVV data in the inner bulge. Steeper near infrared extinction laws such as this lead to best-fit solutions with slightly lower values of both extinction and temperature, not significantly changing our conclusions.}. 
This simple model produced a very good fit for $T=1020$~K, $A_{K_s} = 6.6$ (see upper panel of Fig. \ref{fig:wit01_f6}).
In the lower panel of Fig. \ref{fig:wit01_f6} we illustrate how $\chi^2_\nu$ varies in the parameter space: combinations of temperatures and extinction in the ranges  $720 < T < 2100$~K, $3.6 < A_{K_s} < 10.1$ can produce values of $\chi^2_\nu$ less than four times the minimum value. (As expected, $\chi^2_\nu > 1$ in all models due to systematic uncertainties in the mid-infrared extinction law that become important when extinction is high.)  The best fit extinction value, $A_{K_s} = 6.6$,
is unusually high for an IRDC in the \citet{peretto09} catalogue so within the likely range of $3.6 < A_{K_s} < 10.1$, a lower extinction, slightly lower temperature model 
is perhaps more likely, unless there is some extinction associated with the transient itself.

If we assume the same extinction was present in early March and late August 2010, then the ratio of $K_s$ and $W3$ fluxes from 29 August can be used to calculate a black body temperature. The data indicate significant cooling, e.g. from 1020~K to 713~K (if $A_{K_s} = 6.6$) or from 720~K to 602~K (if $A_{K_s} = 3.6$).
However, the upper limit of $W2>11$ at this epoch corresponds to a fade of over 5~mag at 4.6~$\mu$m, larger than the 4.2 mag fade at 2.15~$\mu$m and the 2.0 mag at 12~$\mu$m. After including this upper limit in $W2$, the SED is not well described by a single temperature black body in late August (see the blue points in Fig.  \ref{fig:wit01_f6}), for any value of extinction. Nonetheless, the relative brightness at 12~$\mu$m suggests a shift to lower temperatures.

Soon after the discovery of the transient in VVV DR1 in early 2012, we obtained deeper and higher resolution follow-up observations on 2012 May 10 using the Infrared Spectrometer And Array Camera (ISAAC, \citealt{moorwood98}) at the ESO Very Large Telescope (ESO programme 289.D-5009). We also obtained {\it Spitzer}/IRAC images on 2012/05/29 (MJD 56076.787), under {\it Spitzer} programme 80259. 
The ISAAC data showed that VVV-WIT-01 had faded to $K_s = 20.34 \pm 0.09$~mag. 
The IRAC data also showed a point source, with $I2$(4.5~$\mu$m) = 15.22 $\pm$ 0.30~mag. (The source is clearly detected but there is uncertainty in the
background level because two much brighter stars are present on either side of VVV-WIT-01 and there are also gradients in the surrounding nebulosity). 
In the $I1$ passband we measure a 3$\sigma$ upper limit, $I1$(3.6~$\mu$m)~$> 16.5$~mag. The $I2$ magnitude can be compared with the earlier WISE $W2$ magnitude, as the passbands are similar. The comparison shows that the variable source has faded by $\sim$9.5~mag at 4.5~$\mu$m over the 2 year interval.
 
\subsection{High energy and neutrino data}
\label{sec:neut}

We note that the ANTARES neutrino project \citep{ageron11} did not detect any non-solar astronomical neutrino sources in the 2007--2012 period (see \url{http://antares.in2p3..fr/publicdata2012.html}). The Super-Kamiokande and SNEWS neutrino projects \citep{fukuda03, antonioli04} also did not detect any events in a 9 month interval prior to March 2012 (M. Vagins and K. Scholberg, private comm.). Similarly, there are no reported detections of $\gamma$-ray or hard X-ray sources that can be connected with this transient in alert data and catalogue data from the {\it Fermi, Swift} and {\it INTEGRAL} satellites \citep{ackermann13, oh18, krivonos12}. 

We moreover obtained targeted X-ray follow up observations of the VVV-WIT-01 field with {\it Swift}/XRT from 21 April to 13 May 2012, totalling 8123~s on source in the 0.3-10~keV passband. No X-ray source was detected, with a 3$\sigma$ upper limit of 3$\times 10^{-14}$~erg cm$^{-2}$ s$^{-1}$. 

\subsection{Radio cm continuum data}
\label{sec:cmdata}

Following the initial announcement of this transient event \citep{minniti12}, the field was observed with the Australia Telescope Compact Array (ATCA) at 2.1~GHz, 5.5~GHz and 9~GHz on 2012 May 21, some two years after the first detections (ATCA programme CX240). The observations used the longest baseline array configuration (6~km) and lasted for a full track, providing good $uv$ coverage and a spatial resolution, $\theta \approx 1\arcsec$~at the two higher frequencies, with a symmetric beam. The phase and gain calibrator was 1613-586. The {\it Miriad} software package was used to reduce the data, using standard procedures to flag and remove bad data.
VVV-WIT-01 was clearly detected at 5.5~GHz and 9~GHz with fluxes of $276 \pm 10~\mu$Jy and $236 \pm 37~\mu$Jy respectively. It was not detected at 2.1~GHz, with a 2$\sigma$ upper limit of approximately 400~$\mu$Jy. A power law fit to the detections and upper limit yields a spectral index $\alpha = -0.15 \pm 0.22$, for $F_{\nu} \propto \nu^{\alpha}$, essentially a flat radio spectrum. 

A second set of ATCA observations at 5.5~GHz and 9~GHz was obtained 7 years later, on 2019 June 24/25, with a duration of almost a full track (ATCA programme C3309). Again the 6~km configuration was used. The phase and gain calibrators observed on this occasion were 1646-50 and 1600-48, the latter being used only at the start of the run before 1646-50 rose high enough. An adjacent, uncatalogued continuum source at 16:10:59.4 -51:49:24.0 was used for self-calibration at 5.5~GHz but this was not possible at 9~GHz. VVV-WIT-01 was no longer detected, with 3$\sigma$ upper limits of approximately 40~$\mu$Jy at both frequencies.

\subsection{Radio mm/submm continuum data}
\label{sec:alma}

The transient was also followed up with the Atacama Large Millimeter/submillimeter Array (ALMA), using the 12~m antennae in the main array. Observations were made in the continuum in ALMA band 3 (2.6--3.6~mm), band 6 (1.1--1.4~mm) and band 7 (0.8--1.1~mm) in 2013 and 2015  (ALMA project codes 2012.1.00463.S and 2013.A.00023.S). These were reduced with the standard data reduction pipeline by observatory staff.
No source was detected at the coordinates of VVV-WIT-01, to 3$\sigma$ upper limits of approximately 0.1 mJy in all 3 bands.  In bands 6 and 7 two sources were detected within the area of SDC~G331.062$-$0.294, both of them offset from VVV-WIT-01 by a few arcseconds. One of these sources is coincident with a likely YSO, VVV~J161053.93-515532.8, that happens to show a $\sim$1~mag infrared outburst in 2015. This is one of the two red stars visible to either side of VVV-WIT-01 in the middle panel of figure 3, both of which are YSO candidates. We give details of VVV~J161053.93-515532.8 in Appendix A. The other ALMA detection has no infrared counterpart and may well be an extragalactic object.

\section{Discussion and searches for additional events}
\label{sec:discussion}

\subsection{Protostellar collision interpretation}
\label{sec:coll}

Owing to the location projected against an IRDC, we must consider the case for a genuine physical association and a pre-main sequence origin.
IRDCs are numerous in the inner Galactic plane but relatively small. Using the equivalent area radii from \citet{peretto16} for IRDCs that passed their far-infrared confirmation test, we derive a covering fraction, $f  = 1.2\%$ within the southern portion of the GLIMPSE-I survey area at $295<l<350^{\circ}$, $|b|<1^{\circ}$, where the IRDC catalogue of \citet{peretto16} was defined. This rises to $f=2.3\%$ if we include IRDCs that did not pass the automated confirmation test. 
These small covering fractions led us to explore the possibility that the transient was a collision between YSOs within the IRDC, SDC G331.062$-$0.294. Collisions, which will often lead to mergers, can cause high amplitude transient events such as the ``mergeburst"  observed in V1309~Sco \citep[arising in a contact eclipsing binary system]{tylenda11} and other likely examples such as V838~Mon and V4332 Sgr \citep{martini99, tylenda06, kochanek14}. As noted earlier, such events are a possible explanation for ``luminous red novae" seen in external galaxies.  

We note that the high amplitude of the event ($\ge$9.5 mag at 4.5~$\mu$m) tends to argue against some form of episodic accretion event in a protostar because infrared variations in the most extreme such events have not exceeded $\sim$6 to 7~mag hitherto, e.g. \citet{audard14}. An apparent protostellar eruption with an unprecedented  mid-infrared amplitude of 8.5~mag has recently been discovered in the public WISE/NEOWISE, GLIMPSE and VVV data (Lucas et al., in prep) but the event has a long plateau and a slow decline over several years, not resembling the rapid decline seen in VVV-WIT-01.

\citet{bally17} provide three examples of explosion remnants within Galactic star formation regions that are likely due to historical collisions between YSOs. The best- studied example is the Becklin-Neugebauer event in OMC-1, where recent data indicate that three YSOs are moving away from a common origin some 500~years ago, at the centre of the explosion remnant \citep{luhman17}. The physics of collisions between protostars, and mergers, is discussed in detail by \citet{bally05}. The topic is interesting in part because mergers are a possible mode of formation for massive stars \citep{bonnell98}, though it is now thought likely that massive stars can form without recourse to this mechanism \citep{kuiper10, baumgardt11, kuiper18}. If collisions in star forming regions do occur, the relatively low space density in such regions would require some explanation other than random motions within the cluster potential, even when gravitational focussing is allowed for \citep{shara02}. Possible explanations might include focussing of YSOs within relatively small volumes due to formation within hub-filament systems of dense gas within molecular clouds \citep{myers09, peretto13, williams18},  ongoing accretion of low angular momentum gas in a star forming cluster \citep{gieles18} or perhaps the unstable dynamical decay of newborn multiple systems \citep{rawiraswattana12, moeckels13, railton14}.

No massive protostar remained visible in the vicinity of VVV-WIT-01 after the transient event so a collision would have to involve low mass YSOs. This is not an obstacle because the great majority of YSOs have low masses and they can appear faint when embedded in an IRDC at a distance of several kpc. IRDCs have typical distances of 1--8~kpc \citep{simon06} and we inferred in section \ref{sec:2} that SDC~G331.062$-$0.294 is probably located in front of the bright HII region IRAS 16067-5145 at $d \approx 7.4$~kpc. The recent 
$^{12}$CO-based molecular cloud catalogue of \citet{miville-desch17} lists at least four massive molecular clouds that appear to overlap the location of VVV-WIT-01, with likely distances ranging from 1.5 to 8.5~kpc.
The distance to SDC~G331.062$-$0.294 can be constrained using blue foreground stars with parallaxes from Gaia. Gaia source 5933412592646831872 is projected in this IRDC and has a likely minimum parallax-based distance of 2.4~kpc in \citet{bailer-jones18}. Source 5933458016223623552, projected in the adjacent, apparently connected IRDC SDC~G331.030$-$0.299 has a likely minimum parallax-based distance of 3.1~kpc (based on a 7$\sigma$ parallax and neglecting the small systematic parallax error in Gaia DR2). Combining these lower limits with the upper limit from the background HII region, we can assume $3.1 < d < 7.4$~kpc for these IRDCs. Unfortunately it was not possible to calculate a red clump giant distance \citep{lopez-corredoira02} because the IRDCs are too small to have a projected population of such stars. The giant branch in the 2MASS and VVV $K_s$ vs $J-K_s$ colour magnitude diagrams for the larger region within 3 arcminutes of VVV-WIT-01 (not shown) shows a fairly smooth increase in extinction with distance at $d>3$~kpc, consistent with the existence of several large molecular clouds along this line of sight.

\subsection{Supernova hypotheses}

The non-detections of neutrinos and $\gamma$-rays (see section \ref{sec:neut}) rule out a highly obscured core collapse supernova on the far side of the Galactic disc. Such events are very luminous $\gamma$-ray and hard X-ray sources; we would also expect to detect the neutrino signal because it is very rare for all three of the neutrino observatories listed in section \ref{sec:neut} to be offline simultaneously. 

These neutrino and high energy radiation checks were necessary because a relatively low luminosity core collapse supernova with peak absolute magnitude $M_{K_s} = -16$, subject to extinction $A_{K_s}=10$~mag (the maximum plausible value, see above) might have apparent magnitude $m_{K_s}=10.5$ if located at $d$=20~kpc from the sun; the infrared data alone would not entirely rule out this possibility because VVV-WIT-01 was discovered after the peak. Similarly, while a core collapse supernova would typically be a far brighter cm continuum radio source than was observed (see section \ref{sec:cmdata}), rare examples with blue supergiant progenitors can be radio-quiet \citep{weiler88}.

A type Ia supernova in the inner Milky Way would be radio-quiet and lack detectable neutrino emission \citep{odrzywolek11}. However, such an event would have been detected in $\gamma$-rays by the {\it Fermi}, {\it Swift} or INTEGRAL telescopes due to strong emission at specific energies associated with the decay of $^{56}$Ni and $^{56}$Co in the ejecta, see \citet{wang19} for a full discussion. For example, in the case of SN2014j at 3.3~Mpc distance, the 847 keV line of $^{56}$Co was detected by INTEGRAL for more than four months \citep{diehl15}.

\subsection{Helium flash hypothesis}

 A few examples of a very late thermal pulse in a post-Asymptotic Giant Branch star have been observed over the past century, e.g. V4334 Sgr (Sakurai's object) and V605 Aql, \citep[see e.g.][]{hajduk05}. In these events, the star returns from the white dwarf cooling track to giant star status but then remains luminous 
 ($L \sim 10^4 L_{\odot}$) for hundreds of years. Such stars can appear almost nova-like at visible wavelengths because the rise in flux takes place over a time of order 
 1 year and subsequent formation of dust in the ejecta can cause a rapid optical fading, as in V605~Aql. The near infrared flux can also fade and re-brighten as
 the ejecta cool and the observed structure changes \citep{hinkle14, evans06, hinkle02, clayton13, kimeswenger00} but the mid-infrared flux remains high for several years, with measured temperatures exceeding 500~K. The data  for VVV-WIT-01 do not seem consistent with a helium flash event because, while some cooling
apparently occurred over the six month interval after first detection,  the source faded by 2~mag at 12~$\mu$m and by much larger amounts at 4.6~$\mu$m and 2.15~$\mu$m (see section \ref{sec:irop}), indicating that the total luminosity probably declined by a large factor during this time. This rapid decline, and the continued decline
from 2010--2012 at 4.6~$\mu$m, contrast with the approximately constant luminosity of V4334~Sgr over several years \citep{hinkle14}. The relative rarity of these events also means that a chance association with an IRDC would be unlikely.

\subsection{Classical Nova interpretation}
\label{sec:nova}

A clear case may be made for a classical nova located behind the IRDC. Classical novae are the commonest type of transient seen in VVV: at least 20 likely examples have been discovered within the VVV dataset, many of them in the ``VVV disc" area at Galactic longitudes $295<l<350^{\circ}$ \citep[e.g.][]{saito15, saito16}  and many others have been independently discovered within the Galactic bulge since 2010, by VVV, WISE or optical surveys such as OGLE and ASAS. Classical novae (and recurrent novae, which are novae that have been observed to erupt more than once) have outburst amplitudes from 7 to 20~mag in the optical \citep[e.g.][]{strope10, osborne15} and a very wide range of decay timescales; in the sample of 95 optical nova light curves presented by \citet{strope10}, the $t_3$ parameter (the time for the outburst to decline by 3~mag from the peak at optical wavelengths) ranges from 3 days to 900 days. 
\citet{strope10} illustrated considerable variety in the morphological features of nova light curves that remains poorly understood. 

In the infrared, only a few well sampled novae have been studied in detail \citep[see e.g.][]{gehrz88, gehrz99, hachisu06, gehrz15}, though fairly well sampled $K$ bandpass light curves are now available for many novae in the SMARTS database (\url{www.astro.sunysb.edu/fwalter/SMARTS/NovaAtlas/}). Many novae have a second peak in the infrared light curve some 30 to 60 days after the initial optical/infrared maximum, due to the condensation of a dust shell within the ejecta that is sometimes optically thick and sometimes optically thin, perhaps due to clumpiness \citep[see e.g.][]{gehrz88, hachisu06, das13, burlak15}. The SED of the dust shell is well fit by emission from a black body at temperatures from $\sim$700 to 1100~K. Examples include Nova Cyg 1978 \citep{gehrz80a} and Nova Ser 1978 \citep{gehrz80b}, whilst in the SMARTS database  the light curves of e.g. V906 Car and V5668 Sgr appear to show a prominent $K$ bandpass dust peak. A large amount of dust production is more common in CO novae, originating in low mass white dwarfs, than ONeMg novae wherein the white dwarf has a typical mass $M>1.2~M_{\odot}$ \citep{gehrz99}.  Our derived temperature near 1000~K for the Planckian emission from VVV-WIT-01 is clearly a close match to a classical nova soon after the epoch of dust condensation.

The absolute visual magnitudes of classical novae at peak brightness typically lie in the range $-10<M_{V}<-5$ \citep{warner87} and $K_s$ magnitudes are similar to $V$ at this time, in the absence of reddening. The absolute $K_s$ magnitude of the second peak after the epoch of dust formation can vary widely depending on the optical depth and temperature of the dust \citep[e.g.][]{hachisu19} but in some of the above examples it is similar to, or only 1 or 2 mag fainter than, the first peak. Adopting our best-fit values of extinction and temperature $A_{K_s}=6.6$, $T \approx 1000~K$ for VVV-WIT-01, the brightest $K_s$ magnitudes in March 2010 would imply a distance modulus $m-M=10.6$ to 15.6, or $d=1.3$ to 13~kpc, if we were observing the initial outburst. If we adopt a lower extinction model more typical of IRDCs, within the likely range of $3.6 < A_{K_s} < 10.1$ (see section \ref{sec:irop}), these distances can rise by up to a factor of three, despite the slightly lower temperature associated with these models. The photometric maximum was not in fact observed so the distance is not well constrained in this interpretation, though it would have to be larger than the minimum distance of 3.1~kpc to the IRDC (see section \ref{sec:coll}). Nonetheless, a classical nova with a prominent dust peak could plausibly be located at a distance of several kpc, i.e. behind the IRDC.

\subsection{The cm continuum test}

The cm continuum data from ATCA provide a valuable test of the protostellar collision and nova hypotheses. Numerous radio shells associated with classical novae have been observed previously. Nearby novae have typical fluxes of the order of a few mJy at 4--9~GHz continuum when observed $\sim$2~years after the event and the radio spectral index at those frequencies is typically approximately flat at that time, consistent with optically thin free-free emission \citep{hjellming79, seaquist08, finzell18}. Optically thin nova shells then fade fairly rapidly at these frequencies as the shell becomes increasingly optically thin, with $S_{\nu} \propto (t-t_0)^k$, where $k$ ranges from -3 to -1 and $t_0$ is the time of the explosion. If VVV-WIT-01 was a classical nova, we would expect the radio emission to have faded
by a factor of at least 4.5, and more likely by one or two orders of magnitude, between the 2012 and 2019 ATCA observations. The ATCA non-detection in 2019 at a 
level an order of magnitude fainter than the 2012 flux is therefore consistent with this interpretation. The flat spectral index and $\sim$0.25~mJy flux level in 2012
are also consistent with a classical nova at a distance of several kpc. 

In the case of a collision between YSOs, numerical simulations of relatively violent collisions (those would produce a transient event of nova-like luminosity) predict between 1\% and 10\% of  the combined stellar mass would be ejected, i.e. a mass of order a few $\times 10^{-2} M_{\odot}$ to a few $\times 10^{-1} M_{\odot}$ for a collision between low mass stars \citep{laycock05, dale06, soker06}. This is supported by an empirical estimate of 1~$M_{\odot}$ ejected, in the case of the massive Becklin-Neugebauer event \citep{bally17}.
The ejection velocities would be lower than in a classical nova, of order $10^2$~km/s (i.e. a little less than the escape velocity): this is in line with the ejection velocities observed in V1309~Sco, V4332~Sgr and V838~Mon \citep[][and references therein]{tylenda06, kaminski09, kaminski15}.  
Classical novae typically eject a few $\times 10^{-4}~M_{\odot}$, e.g. \citet{gehrz99}, at speeds of order $10^3$~km/s. Since the ejected mass from a stellar collision is likely to be two or three orders of magnitude greater and the velocity is likely to be almost an order of magnitude less, we would expect the collision ejecta to remain optically thick in the 4~cm continuum for $\sim$2 orders of magnitude longer than the 2 year timescale that is characteristic of novae. Free-free emission from optically thick ejecta tends to slowly brighten over time as the surface area increases, as is observed in the early stages of nova shell expansion, so in such an event we would have expected the radio continuum flux to be brighter in the 2019 ATCA observation than in 2012. The non-detection in 2019 therefore argues fairly strongly against the collision hypothesis.

\subsection{Searches for additional very red transients}
\label{sec:addtran}

In order to further test our two main hypotheses, we searched the current working database of VVV/VVVX 2010--2018 light curves covering the original 560~deg$^2$ area (see section \ref{sec:2}) to determine how many transient events have occurred and whether there have been any other exceptionally red transients in star forming regions.
We detected 59 transient events with amplitudes over 4~mag, 12 of which occurred within the boundary of the GLIMPSE-I survey at $295<l<350^{\circ}$, $-1<b<1^{\circ}$ where the IRDC catalogue of \citet{peretto09} is defined. None of these events were projected in IRDCs and none have SEDs as red as VVV-WIT-01.
Some events that were detected by WISE, e.g. VVV-NOV-005 \citep{saito15}, have colours in the range $1<W1-W2<2$, consistent with emission from hot dust.
Most of these 59 transients have been previously identified as novae or nova candidates but some are new: their infrared light curves will be the subject of a separate paper. 

With 12 events over the 2010--2018 interval in the GLIMPSE~I survey area, the IRDC covering fractions, $f$, noted in section \ref{sec:coll} imply that the chance of observing at least one in an IRDC is $p=1-(1-f)^{12}=0.13$ (including only confirmed IRDCs) or $p=0.24$ (including all IRDC candidates). In view of this result, the nova hypothesis appears very plausible. It was merely surprising, at the time, to detect such a chance projection amongst the first VVV transients near the start of the survey. A separate search for transients in IRDCs was conducted with the WISE/NEOWISE time series data from 2010 and 2014--2017. We constructed a variable star and transient source catalogue for all sources within 2\arcmin~of the 7139 confirmed IRDCs listed in \citet{peretto16}.
This search did not find any additional transients, though some high amplitude variable sources were detected (Lucas et al., in prep).

\subsection{Upper limit on the protostellar collision rate}

The ATCA data and the incidence of VVV transients have led us to conclude that VVV-WIT-01 was very probably a classical nova behind an IRDC. The main caveats are that our empirical knowledge of stellar mergers is limited to a few probable events and the present generation of numerical simulations is unlikely to be the final word. We can use the non-detection of stellar mergers in VVV to place a limit on the incidence of luminous collisions between YSOs. Our search of the 2010--2018 VVV data spanned $\sim$8.5 years and covered an area of sky that encompasses $\sim 10^{11}$ stars (assuming the Milky Way contains at least twice that number of stars). Adopting a constant star formation rate (for simplicity) and typical stellar lifetimes in excess of $10^{10}$~yr, a fraction of order 10$^{-4}$, i.e. 10$^{7}$ stars, have ages $\le$1~Myr. Our detection efficiency is $\approx$0.5, given that events occurring in the southern summer may have been missed by our search. The non-detection in 8.5 years then implies that the rate is probably $\le$0.2 collisions per year. For 10$^{7}$ YSOs with ages $\le 1$~Myr, likely to still be in a crowded and dynamically unstable environment, a limit of 0.2 collisions per year implies a 1 in $5\times10^7$ chance of a collision per year for individual YSOs, or 1 in 50 per Myr per YSO. This upper limit is at a level that begins to be useful, confirming that, while such collisions are rare, there may be numerous such events within the lifetime of massive pre-main sequence clusters of several thousands stars, such as the Orion Nebula Cluster. A careful search of datasets such as VVV, GLIMPSE, WISE and the United Kingdom Infrared Deep Sky Survey (UKIDSS) may yield more examples of the remnants of pre-main sequence mergebursts, to add to the three listed by \citet{bally05}.

\section{Conclusions}

The VVV-WIT-01 transient event was uniquely red amongst transients detected in the VVV dataset thus far. It was also unique in its projected location in an IRDC.
The contemporaneous timing of observations with VISTA and WISE in March 2010 was fortunate but the sparse sampling of the light curve and lack of a spectrum
make a definitive classification difficult. Nonetheless, the absence of $\gamma$-ray and neutrino emission allows us to rule out a Galactic supernova
and the very high amplitude and steep decline in flux make it unlikely that this was an episodic accretion event in a YSO. The steep decline in mid-infrared flux also appears to rule out a Very Late Thermal Pulse in a post-AGB star.

The hypothesis of a highly obscured classical nova with a bright dust peak appears to fit all the available data. Whilst it was initially surprising that such an event should be projected in an IRDC in the first year of the survey, VVV subsequently detected several dozen transients in the 2010--2018 interval, including 12 in the {\it Spitzer}/GLIMPSE region where the \citet{peretto09} IRDC catalogue was defined. Consequently, the chance of detecting a single such event over the course of the survey is calculated as between 13\% and 24\%, not a very unlikely occurrence. 

The protostellar collision/mergeburst hypothesis is the most interesting one that we have considered, given that the remnants of a few such events involving pre-main sequence stars have been observed in the Milky Way \citep{bally05}. The rapidly fading 4~cm continuum flux detected by ATCA appears to be inconsistent with this interpretation, causing us to favour the nova hypothesis. A cautionary note remains that our prediction regarding the radio evolution of a mergeburst rests on theoretical simulations in a field that is still maturing. We should remain open to the possibility that the transient was a new type of event. Any detections of additional very red transients in star forming regions would require us to reconsider the classical nova interpretation.

\section*{Acknowledgements}

We gratefully acknowledge data from the ESO Public Survey program ID 179.B-2002 taken with the VISTA telescope, and products from the Cambridge Astronomical Survey Unit (CASU). We thank the referee, Nye Evans, for a helpful and constructive report. We also thank M. Vagins and K. Scholberg for checking their neutrino
data during the relevant 2009-2010 time period. This publication makes use of data products from the WISE satellite, which is a joint project of the University of California, Los Angeles, and the Jet Propulsion Laboratory/California Institute of Technology, funded by the National Aeronautics and Space Administration (NASA). This research has made use of NASAÕs Astrophysics Data System Bibliographic Services and the SIMBAD database operated at CDS, Strasbourg, France. P.W.L. acknowledges support by STFC Consolidated Grants ST/R00905/1, ST/M001008/1 and ST/J001333/1 and the STFC PATT grant 
ST/R00126X/1. 
D.M. acknowledges support from the FONDECYT Regular grant No. 1170121,  the BASAL Center for Astrophysics and Associated Technologies (CATA) through grant AFB170002, and the Ministry for the Economy, Development and Tourism, Programa Iniciativa Cientifica Milenio grant IC120009, awarded to the Millennium Institute of Astrophysics (MAS). D.L.K. was supported by NSF grant AST-1816492. N.M. acknowledges financial support from ASI-INAF contract n.2017-14-H.0. M.H. acknowledges financial support from the Comit\'{e} Mixto ESO-Gobierno de Chile. Support for M.C. is provided by Proyecto Basal AFB-170002; by the Ministry for the Economy,
Development, and Tourism's Millennium Science Initiative through grant IC\,120009, awarded to the Millennium Institute of Astrophysics (MAS); and by FONDECYT project \#1171273.



\bibliographystyle{mnras}
\bibliography{witbib} 



\appendix
\section{A submm-bright variable YSO in SDC~G331.062$-$0.294}

\begin{figure}
	\includegraphics[width=\columnwidth]{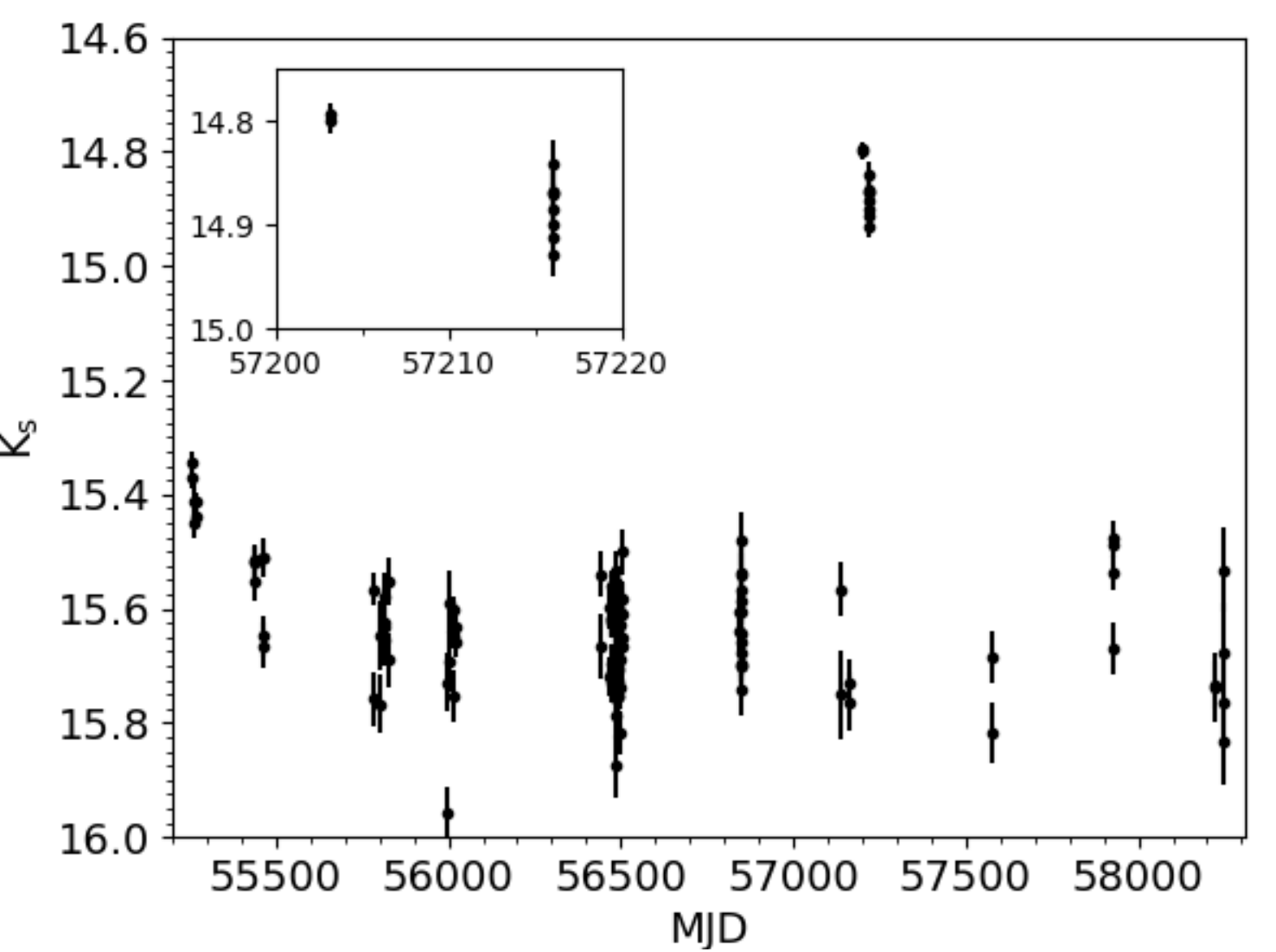}
   \caption{VVV $K_s$ light curve for the submm-bright source VVV~J161053.93-515532.8. A jump of $\sim$0.9~mag occurred in mid-2015, lasting for at least
   13 days. The flux had returned to the pre-outburst level in data taken 12 months later. The inset panel shows an expanded view of the data during the outburst.}
    \label{fig: mmvar}
\end{figure}

As noted in section \ref{sec:alma}, the ALMA follow up observations of VVV-WIT-01 did not detect the transient but a submm/mm continuum source was detected 4.5\arcsec~to 
the east. The submm flux was $0.62 \pm 0.04$ mJy in band 6 (230~GHz) and $0.86 \pm 0.06$ mJy in band 7 (336.5 GHz). It was not detected in band 3 (100 GHz) to a 3$\sigma$ limit of 0.14~mJy. The band 6 observations were obtained on 2015 August 14 and 2015 September 4 and the band 7 observations were on 2015 September 25. The beam sizes were 0.5\arcsec, 0.3\arcsec~and 0.25\arcsec, in band 3, band 6 and band 7, respectively.
The ALMA source is coincident (within 0.1\arcsec) with the red VVV point source, VVV~J161053.93-515532.8, that is visible about 5\arcsec~east of VVV-WIT-01 in the middle panel of Fig. 3. This red star has mean $K_s =15.39$~mag, $H-K_s$=1.66, $J-H$$>$3. The VVV/VVVX $K_s$ light curve is shown in Fig. \ref{fig: mmvar}.
Our {\it Spitzer}/IRAC data (see section \ref{sec:irop}) give magnitudes $I2 = 12.39$, $I1-I2 = 0.71$.

The infrared colours and fluxes are typical for a low mass class I or class II YSO \citep{megeath04, gutermuth09} at a distance of a few kpc and in view of the projected location in an IRDC it is likely that VVV~J161053.93-515532.8 is indeed a YSO. The $K_s$ light curve clearly shows an infrared outburst of 0.9~mag amplitude in the data taken on 30 June and 13 July 2015 (MJDs 57203 and 57216). This further supports a YSO identification, in view of the fact that high amplitude variability is fairly common in YSOs \citep{cp17a}. The infrared burst faded by 0.08 mag over the 13 days between the two dates of observation. The infrared flux was otherwise fairly constant from 2012--2018 and was observed at the quiescent level on 2015 May 19 and 2016 July 5. A slow 0.25~mag decline was seen from 2010 to 2011. 

The ALMA continuum detections can be attributed to emission by dust in the cool outer regions of a circumstellar disc. This would be consistent with the non-detection
in band 3, due to the decline in dust emissivity with increasing wavelength. We can estimate the dust mass with the equation
$M_{dust} = (F_{\nu} d^2) / (\kappa_\nu B_{\nu}(T))$ \citep{hildebrand83}. Adopting $T$ = 20~K and opacity $\kappa = 10 (\nu/1000)^{\beta}$~GHz cm$^2$ g$^{-1}$ and $\beta=1$ \citep{beckwith90} yields a mass $M_{dust} = 5 \times 10^{-4} (d/3$~kpc)$^2$~M$_{\sun}$. Adopting the conventional gas to dust ratio of 100:1, this implies $M_{disc} = 0.05 (d/3$~kpc)$^2$~M$_{\sun}$. Recalling that the distance to SDC~G331.062$-$0.294 is likely to be between 3.1 and 7.4~kpc, this indicates that the YSO hosts a relatively massive disc, or perhaps an unusually warm one. It is certainly possible that the infrared outburst was ongoing at the time of the ALMA detections in August and September 2015, resulting in heating of the outer disc \citep{johnstone13}.


\bsp	
\label{lastpage}
\end{document}